# Multicanonical multigrid Monte Carlo method and effective autocorrelation time[*]

W. Janke[a][†] and T. Sauer[b]

[a]Institut für Physik, Johannes Gutenberg-Universität Mainz, 55099 Mainz, Germany

[b]Institut für Theoretische Physik, Freie Universität Berlin, 14195 Berlin, Germany

We report tests of the recently proposed multicanonical multigrid Monte Carlo method for the two-dimensional $\Phi^4$ field theory. Defining an effective autocorrelation time we obtain real time improvement factors of about one order of magnitude compared with standard multicanonical simulations.

## 1. INTRODUCTION

At first-order phase transitions [1] standard Monte Carlo simulations in the canonical ensemble exhibit a supercritical slowing down. Here extremely large autocorrelation times are caused by strongly suppressed transitions between coexisting phases which, on finite periodic lattices, can only proceed via mixed phase configurations containing two interfaces. Since the probability of such configurations is suppressed by a factor $\exp(-2\sigma L^{d-1})$, where $\sigma$ is the interface tension and $L^{d-1}$ the cross-section of the system, the autocorrelation times in the simulation grow exponentially with the size of the system, $\tau \propto \exp(2\sigma L^{d-1})$. A way to overcome this problem, known as multicanonical sampling [2], is to simulate an auxiliary distribution in which the mixed phase configurations have the same weight as the pure phases and to compute canonical expectations by reweighting [3]. While this does reduce the supercritical slowing down to a power-law behaviour the remaining slowing down problem is still severe. In fact, in most cases it is even worse than for standard (e.g., Metropolis or heat-bath) Monte Carlo simulations of critical phenomena. For these latter applications, on the other hand, multigrid techniques [4–7] have been shown to greatly reduce or even completely eliminate critical slowing down. Here collective updates on different length scales are performed by visiting various coarsened grids in a systematic, recursively defined way. For a further reduction of autocorrelations both approaches may easily be combined and give a much better performance than each component alone [8].

## 2. SIMULATION

We studied the $\Phi^4$ lattice field theory in $d=2$ dimensions defined by the partition function

$$Z = \prod_i^{L^d} \left[ \int d\Phi_i \right] \exp\left( -\sum_{i=1}^{L^d} \left( \frac{(\vec{\nabla}\Phi_i)^2}{2} - \frac{\mu^2}{2}\Phi_i^2 + g\Phi_i^4 \right) \right) \quad (1)$$

with $\mu^2, g > 0$. Here reflection symmetry is spontaneously broken for all $\mu^2 > \mu_c^2(g) > 0$ as $L \to \infty$. Consequently, if a term $h\sum_i \Phi_i$ is added to the energy, the system exhibits first-order phase transitions driven by the field $h$.

For the multicanonical sampling the reweighting factor is denoted by $w^{-1}(m) \equiv \exp(-f(m))$, where $m = \sum_i \Phi_i / V$ is the average field. Canonical expectation values $\langle \mathcal{O} \rangle_{\text{can}}$ of an observable $\mathcal{O}$ are then obtained by the basic reweighting formula $\langle \mathcal{O} \rangle_{\text{can}} = \langle w\mathcal{O} \rangle / \langle w \rangle$, where $\langle \ldots \rangle$ on the r.h.s. are multicanonical expectation values. To update field values with, say, Metropolis moves, $\Phi_i \to \Phi_i + \Delta\Phi_i$, the decision of acceptance is now based on the value of $\Delta E + f(m + \Delta\Phi_i/V) - f(m)$ with $\Delta E$ being the canonical energy difference.

[*]To appear in the Proceedings of LATTICE '93, Dallas, USA, October 1993. Talk presented by T. Sauer.
[†]W.J. thanks the Deutsche Forschungsgemeinschaft for a Heisenberg fellowship.

For the multigrid Monte Carlo we use the piece-wise constant interpolation scheme which amounts, in the equivalent unigrid viewpoint, to proposing moves for blocks of $1, 2^d, 4^d, \ldots, V = L^d = 2^{nd}$ adjacent variables in conjunction. In a canonical simulation a multigrid update at level $k$ thus consists in considering a common move $\Delta\Phi$ for all $2^{kd}$ variables of one block, $\Phi_i \longrightarrow \Phi_i + \Delta\Phi$, $i \in$ block. For the sequence of length scales $2^k, k = 0, \ldots, n$ we use the W-cycle.

For the multicanonical multigrid simulation the modifications are now rather simple. Since at level $k$ the proposed move would change the average field by $2^{kd}\Delta\Phi/V$, the decision of acceptance is now to be based on the value of $\Delta E + f(m + 2^{kd}\Delta\Phi/V) - f(m)$, where $\Delta E$ is to be computed as in the canonical case. While this modification is obvious from the *uni*grid viewpoint, it should be stressed that in the recursive *multi*grid formulation the multicanonical modification is precisely the same.

For a fair comparison with canonical simulations, we *define* for multicanonical simulations an effective autocorrelation time $\tau^{\text{eff}}$ by the standard error formula for $N_m$ correlated (multicanonical) measurements, $\epsilon^2 = \sigma_{\text{can}}^2 2\tau^{\text{eff}}/N_m$, where $\sigma_{\text{can}}^2 = \langle \mathcal{O}_i^2 \rangle_{\text{can}} - \langle \mathcal{O}_i \rangle_{\text{can}}^2$ is the variance of the *canonical* distribution of single measurements. Here $\epsilon^2 = \sigma_{\hat{\mathcal{O}}}^2 = \langle \hat{\mathcal{O}}^2 \rangle - \langle \hat{\mathcal{O}} \rangle^2$ is the variance of the (weakly biased) estimator $\hat{\mathcal{O}} = \sum_1^{N_m} w(m_i)\mathcal{O}_i / \sum_1^{N_m} w(m_i) \equiv \overline{w_i\mathcal{O}_i}/\overline{w_i}$ for $\langle \mathcal{O} \rangle_{\text{can}}$. This variance can be estimated by jack-knife blocking procedures, or by applying standard error propagation to the variance of $\hat{\mathcal{O}}$, which involves the (multicanonical) variances and covariances of $w_i\mathcal{O}_i$ and $w_i$, and the three associated autocorrelation times $\tau_{\mathcal{O};\mathcal{O}} \equiv \tau_{\mathcal{O}}$, $\tau_{w\mathcal{O};w\mathcal{O}} \equiv \tau_{w\mathcal{O}}$, and $\tau_{w\mathcal{O};\mathcal{O}} = \tau_{\mathcal{O};w\mathcal{O}}$ [8]. By symmetry, for $\mathcal{O} = m$ this simplifies to

$$\epsilon^2 = \frac{\langle w_i m_i; w_i m_i \rangle}{\langle w_i \rangle^2} \frac{2\tau_{wm}}{N_m} \equiv \sigma_{\text{muca}}^2 \frac{2\tau_{wm}}{N_m}, \qquad (2)$$

where $\langle x; y \rangle \equiv \langle xy \rangle - \langle x \rangle \langle y \rangle$ and $\tau_{x;y} = 1/2 + \sum_k \langle x_0; y_k \rangle / \langle x_0; y_0 \rangle$ is the integrated autocorrelation time of multicanonical measurements. In this way properties of the multicanonical distribution (given by $\sigma_{\text{muca}}^2$) are disentangled from properties of the update algorithm (given by $\tau_{wm}$). Note that in $\tau^{\text{eff}} = (\sigma_{\text{muca}}^2/\sigma_{\text{can}}^2)\tau_{wm}$, it is the integrated autocorrelation time of $w(m)m$ that enters and not the exponential autocorrelation time $\tau_m^{(0)}$, as previously investigated [9].

## 3. RESULTS

In our studies of model (1) we investigated the first-order phase transition between the two ordered phases at the points $g = 0.25$ and $\mu^2 = 1.30$, 1.35, and 1.40 which are sufficiently far away from the critical point at $\mu_c^2 = 1.265(5)$ [10] to display the typical behavior already on quite small lattices. A sensitive measure of the strength of the transition is the interface tension $\sigma_{oo}$ between the + and − phase. For $\mu^2 = 1.30$ and $L \to \infty$ this turns out [8] to be $\sigma_{oo} = 0.03443(47)$ which is comparable to the analytical result [11] of $\sigma_{od} = 0.03355\ldots$ for the order-disorder interface tension in the two-dimensional 9-state Potts model. For $\mu^2 = 1.35$ we find $\sigma_{oo} = 0.09785(60)$ and for $\mu^2 = 1.40$ the interface tension is $\sigma_{oo} = 0.16577(73)$ [8].

We performed multicanonical simulations using the Metropolis update and the W-cycle without post-sweeps for lattices of size $V = L^2$ with $L = 8, 16$ and 32. With the multigrid algorithm we also studied lattices of size $L = 64$. After thermalization, each time series contains a total of $10^6$ measurements taken every $n_e$th sweep. The number of sweeps between measurements, $n_e$, was adjusted in such a way that in each simulation the length of each time series is at least $20,000\,\tau_{wm}$.

In Table 1 we give for both update algorithms the various autocorrelation times of the magnetization $m$ which reflects most directly the tunneling process. We see that $\tau_m$ and $\tau_m^{(0)}$ agree well with each other, showing that the corresponding autocorrelation function can be approximated by a single exponential. For $wm$ we obtain values for $\tau^{(0)}$ that are consistent with those for $m$ within error bars. The integrated autocorrelation times, however, are significantly lower, implying that the autocorrelation function is composed of many different modes. We also observe that the difference between $\tau_{wm}$ and $\tau^{\text{eff}}$ can be quite appreciable. From $L = 8$ to $L = 64$ the ratio $\tau^{\text{eff}}/\tau_{wm} = \sigma_{\text{muca}}^2/\sigma_{\text{can}}^2$ varies from about 1.9 to 4.6, reflecting the varying probability distribution shapes with increasing $L$. By fitting $\tau^{\text{eff}}$ to a power law, $\tau^{\text{eff}} \propto L^z$, we obtain for both update algorithms an exponent of about $z \approx 2.3, 2.7$, and 3.0 for $\mu^2 = 1.30, 1.35$, and 1.40; see Fig.1.

Table 1
Autocorrelation times in units of sweeps resp. cycles for the Metropolis (M) or multigrid W-cycle (W) update in multicanonical simulations of the model (1) with $g = 0.25$ and $\mu^2 = 1.30$.

|  | $L = 8$ |  | $L = 16$ |  | $L = 32$ |  | $L = 64$ |
| --- | --- | --- | --- | --- | --- | --- | --- |
|  | M | W | M | W | M | W | W |
| $\tau_m^{(0)}$ | 212(12) | 11.30(32) | 668(23) | 37.2(2.0) | 3120(200) | 148(11) | 746(62) |
| $\tau_m$ | 204.4(4.0) | 10.88(12) | 690(11) | 34.69(76) | 2984(63) | 150.0(4.0) | 758(37) |
| $\tau_{wm}^{(0)}$ | 209(12) | 11.34(33) | 655(31) | 36.9(2.0) | 2880(190) | 146(13) | 600(120) |
| $\tau_{wm}$ | 171.1(3.4) | 9.82(11) | 509.8(8.9) | 27.58(59) | 1840(40) | 96.6(2.4) | 374(23) |
| $\tau^{\text{eff}}$ | 322.7(6.1) | 18.51(20) | 1258(21) | 67.4(1.3) | 6050(120) | 321.9(7.6) | 1724(86) |

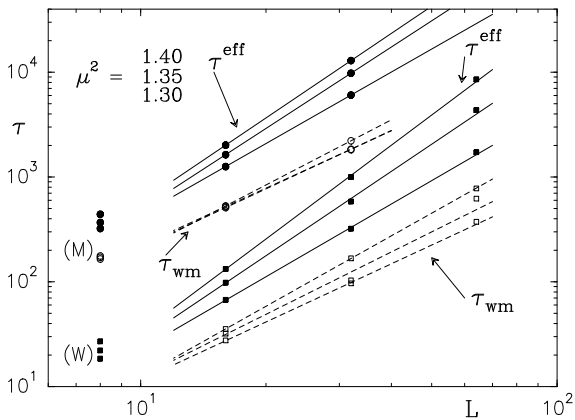

Figure 1: Autocorrelation times as a function of lattice size $L$ for $g = 0.25$.

## 4. CONCLUSIONS

The multigrid update enhances the performance of the multicanonical simulation by reducing the overall scale but it does not affect the exponent $z$. For the W-cycle the autocorrelation times are reduced by a roughly constant factor of about 20 as compared with the Metropolis algorithm. Taking into account that a W-cycle requires more elementary operations than a Metropolis sweep [4] we obtain a *real time* improvement factor of about 10 with our implementation on a CRAY Y-MP.